\documentclass[11pt,twoside]{article}

\usepackage{asp2006}
\usepackage{graphicx}

\def\msun{M$_{\odot}$}
\newcommand{\mum}{\,$\mu$m}

\begin{document}
\title{Spitzer-IRAC GLIMPSE of High Mass Protostellar Objects}  
\author{M. S. N. Kumar \& J. M. C. Grave}  
\affil{Centro de Astrofisica da Universidade do Porto, Rua das Estrelas, s/n, 4150-762, Porto, PORTUGAL}    

\begin{abstract}
The Spitzer-GLIMPSE point source catalog and images have been used to
study a sample of 381 massive protostellar candidates. IRAC-Point
source photometry was used to analyse colours, magnitudes and spectral
indicies of the infrared counterparts (IRCs) of massive protostellar
objects and a bonafide sample of 50 point sources was
obtained. Spectral energy distributions (SEDs) of these 50 sources was
extended to the near-infrared and millimeter range by using 2MASS and
millimeter data from the literature. An online SED fitter tool based
on Monte-Carlo radiative transfer of an accretion model involving
star,disk and envelope was used to fit the SEDs of the 50 sources. The
IRCs to massive protostellar objects are found to successfully imitate
the SEDs of evolutionary phases similar to low mass star
formation. Envelope accretion, rather than disk accretion is found to
be dominant in building the most massive stars. Unresolved centimeter
continuum emission is associated with 27 IRCs classified as massive
protostars suggesting that ionised accretion flows may play an
important role along with the molecular component. The morphology of
the infrared nebulae surrounding the IRCs have an unusual resemblance
to the morphologies of ultra-compact HII regions suggesting that these
infrared nebulae are possible precursors to the UCHII regions.
\end{abstract}

\section{Introduction}   

In recent years, extensive observational data in the infrared and
millimeter bands have led to the belief that the formation of massive
stars mostly involves accretion phenomenon similar to the formation of
low mass stars \citep[e.g.][]{whit05}. Whether the massive protostars are
scaled up versions of low mass protostars or if they go through
continuuing accretion on to zero-age main sequence stars is an issue
under debate \citep[e.g: see the review by][]{zy07}. Addressing these
issues through direct observations requires angular resolutions of an
order of magnitude better than what is currently possible in the
infrared and millimeter range. The recently made available GLIMPSE
survey already provides a breakthrough in the spatial scales and
sensitivity which is much higher than the upto now popular MSX and
IRAS surveys. The Spitzer-IRAC point spread function of $\sim$2\arcsec\,
in the 3.6-8\mum\, bands are comparable to that of 2MASS in the J,H,K
bands which allows us to probe scales of a few thousand AU at
mean distances of 3-5\,kpc where high mass protostellar objects (HMPOs)
are commonly found. Similarly 1\arcsec\, resolution
observations in the millimeter bands made possible by the Plateau de Bure
interferometer or the Sub Millimeter Array has unveiled disks and
toroids around several HMPOs \citep[e.g.][]{bel05,beu07}.

A decade long effort, that used systematic selection criteria
involving constraints on the IRAS colours and other signposts of
massive star formation such as masers and outflows, has resulted in
four HMPO candidate samples, covering the northern and southern
hemispheres of the sky
\citep{mol96,sri02,fon02,faun04}(hereafter Mol96, Sri02, Fon02,
Faun04). These surveys have provided a sample of $\sim$ 500 objects
and a major fraction of the sample has been investigated extensively
to evaluate the dense gas, dust continuum, ionised emission and maser
associations. Using the 2MASS point source catalog, \citet{kumar06}
looked for clustering around the HMPOs in the northern hemisphere and
found 54 embedded clusters and several near-infrared counterparts to
these objects. In this work, we extend such as study to a more complete
sample of HMPOs using the GLIMPSE survey data.

\section{Infrared counterparts to HMPOs} 

Of the 500 targets from the Mol96, Sri02, Fon02 and Faun04, 381
targets were covered by the GLIMPSE fields for which point source
photometry and images were extracted. Point sources within a 100\arcsec\,
radius centered on each target were used for colour, magnitudes and
spectral index analysis whereas images with a size of 300\arcsec\, were
extracted for examining nebulosities and clustering. The distance to
each target was taken as listed in the original references and a mean
distance of 5\,kpc was assumed for sources when an estimate was not
available. The point sources were placed on a [3.6-4.5] vs [5.6-8.0]
colour-colour diagram and separated into the evolutionary stages of
Class I, II and III based on the definitions of
\citet{robitaille06}(hereafter RWIWD06). The fluxes in the four IRAC
bands namely 3.6, 4.5, 5.8 and 8.0 $\mu$m bands were fitted with a
simple least squares linear fit to obtain the spectral index $\alpha$
in these bands. Since the interest is not only to finding redenned
objects but also luminous objects, we defined a parameter called
alpha-magnitude product AM = -M$_{8\mu m}\times$($\alpha$+6)/10 where
M$_{8\mu m}$ is the 8.0$\mu$m\, absolute magnitude of the source and
$\alpha$ the observed IRAC spectral index. The constants 6 and 10 in the
above equation were chosen arbitrarily to effectively separate the
high $\alpha$ sources from the rest. For more details see \citet{kumar07a}.
 
In Fig.~1a we show a [3.6-4.5] vs [5.6-8.0] colour-colour diagram for
all the sources detected from all target lists. The main concentration
of points at (0,0) represents photospheres and higher values on both
the x and y axes represent young stellar objects (YSOs) considered to
be at different evolutionary stages. The regions representative of the
evolutionary stages such as I, and II based on the 2D radiative
transfer model data of RWIWD06 are marked. The diamonds represent
candidate massive protostars which are those points lying to the right
side of the vertical dashed line in Fig1b. Fig.~1b shows a AM product
vs M$_{8\mu m}$ (absoulte magnitude) for point sources from all the
targets. The solid curves represent 20\msun and 8\msun\, class I
object models for all inclinations computed by RWIWD06. Fig.~1c shows
the histogram of the IRAC spectral indicies $\alpha$ for all observed
point sources. The histograms of the IRAC $\alpha$ for YSOs in the
Orion cloud and stars$+$YSOs in the IC348 region are shown for
comparision. The colour and magnitude analysis shows that several
point sources observed close to the HMPO targets display infrared
excess emission in the IRAC bands that imitate the colours of YSO's in
the Class I, II, III phases, and a good fraction of these are also
luminous. Many point sources that are saturated in the IRAC bands may
actually qualify as better IRCs. 37 saturated sources have been found
to coincide well with the HMPO target positions in our sample.

The colours and magnitudes analysis presented in Fig.~1 shows that:
a) the point sources in the HMPO fields have unusually high $\alpha$
values ($\alpha \ge$5), even higher than many of the deeply embedded
Class I sources in Orion, showing that IRCs are indeed deeply embedded
sources inside very dense cores and, b) a significant fraction of
these IRCs are also luminous and represent 8-20\,M$_{\odot}$ Class I
type objects when compared with accretion models.
 
\begin{figure} 
  
  \includegraphics[width=13cm]{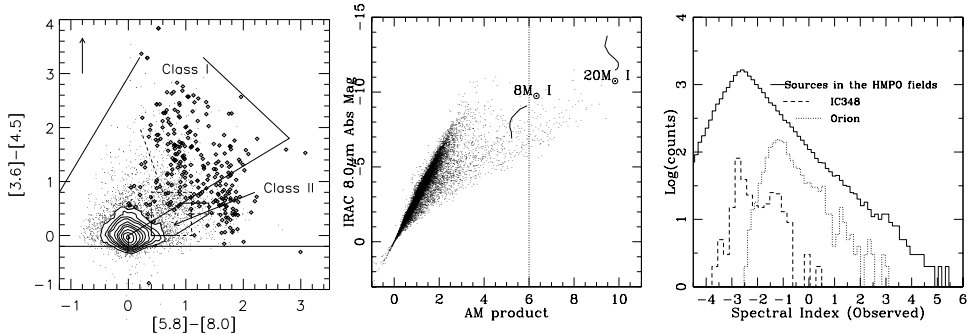} 
   \caption{Histogram of the observed spectral indices ($\alpha$) for
   sources associated with candidate massive protostellar
   candidates. The dotted curve shows a similar histogram for the IRAC
   YSOs in Orion A cloud and the dashed curve shows
   the distribution of sources in IC348.}

  \end{figure} 

A total of 79 sources with AM$\ge$6 were classified as luminous IRCs
to HMPO targets with photometry available in all four IRAC bands. Of
these, 11, 27, 23 and 25 are from the samples of Mol96, Sri02, Fon02,
and Faun04 respectively and does not include the 37 saturated
sources. These luminous IRCs are centred on the IRAS/mm peaks in the
target fields with a nominal spread consistent with beam sizes and
positional uncertainities. Some of these luminous IRCs can be reddened
photospheres (particularly those with [5.8-8.0]$<$0.4 colours) and
some are probably evolved protostars; therefore, a complete SED
analysis described in Sec.~4 was made.

\section{Infrared Nebulae}

The IRAC images of the HMPO targets reveal compact nebulae
(10\arcsec-60\arcsec angular sizes) around several sources. The
nebulae are found to be brightest in the 8\mum\, band and mostly
invisible in the 3.6\mum\, IRAC band. We have used the 4.5, 5.8 and
8.0 \mum\, band images coded as blue, green, and red, respectively to
generate composite false colour images for the observed nebulae. The
analysis of the images are described by \citet{kumar07a} and are also
available online\footnote{The colour images can be accessed at
http://www.astro.up.pt/$\sim$nanda/hmpo/}. The infrared nebulae
repeatedly display cometary (e.g.: I18437-0216), core-halo (e.g:
I18337-0743, I19403+2258), shell-like (e.g.: I19198+1423) and bipolar
(e.g.:I18530+0215, I19213+1723) morphologies which are similar to the
morphologies of UCHII regions \citep{kurtz94} and bipolar outflows
. Some images display only a single well-defined nebula, whereas
others show a group of compact nebulae. A histogram of the projected
sizes of these nebulae displays a distribution in the range
0.1-1.0\,pc with a mean value of 0.5\,pc. These dimensions are similar
to or smaller than the size of dense cores traced by the 1.3\,mm
continuum maps \citep{beu02}. The 8.0\mum\, band is dominated by PAH
emission, which is known to be a tracer of radiation
temperature. Ionising radiation from young massive stars, that may not
yet be strong enough to produce a significant ionised region may
therefore be traced by these infrared nebulae. Indeed a recent
investigation has shown that the underlying structure of the ISM in
such nebulae can possibly be inferred using the morphology of the
nebulae at various density regimes and ionising fluxes
\citep{heitsch07}. Therefore, the morphology of the nebulae found here
may well indicate the morphology in which ionising radiation is
escaping from the underlying set up of physical structures close to
the star. Recently \citet{churchwell06} used the GLIMPSE images to
identify bubbles around OB stars in the Galaxy and argue that the
smaller bubbles around several B stars are those produced by
relatively soft radiation that fails to produce significant HII
regions. The infrared nebulae around the HMPOs may well represent such
bubbles or could be simple reflection nebulae due to an evolved
generation of B stars.

\section{Radiative Transfer Modelling of the SEDs}

As described in Sec.~2, the high AM product sources are our best
candidates of luminous IRCs to the HMPO targets. A bonafide list of 50
IRCs was made from the 79 high AM IRCs by using the constraint that
the high AM product IRC was found within 2\arcsec of the observed
mm/submm peak. We then extended the SEDs of this bonafide sample
to the near-infrared and millimetre wavelength range by using the
2MASS point source catalog, millimeter observations from \citet{beu02}
and the submillimeter data from SCUBA \citep{wil04} and SIMBA cameras
\citep{bel06}. Data with the highest available angular resolution
was used for the SEDs. In two cases, interferometric observations in
the mm and submm range were available that was utilised. The IRAC and
2MASS fluxes are restricted to their PSF sizes of less than 2\arcsec
and typical mm/submm beam widths are $\sim$6\arcsec. The resulting
SEDs were fitted with an online SED fitter tool developed and tested
on low mass YSOs by \citet{rob07}.

\begin{figure} 
  
  \includegraphics[width=13cm]{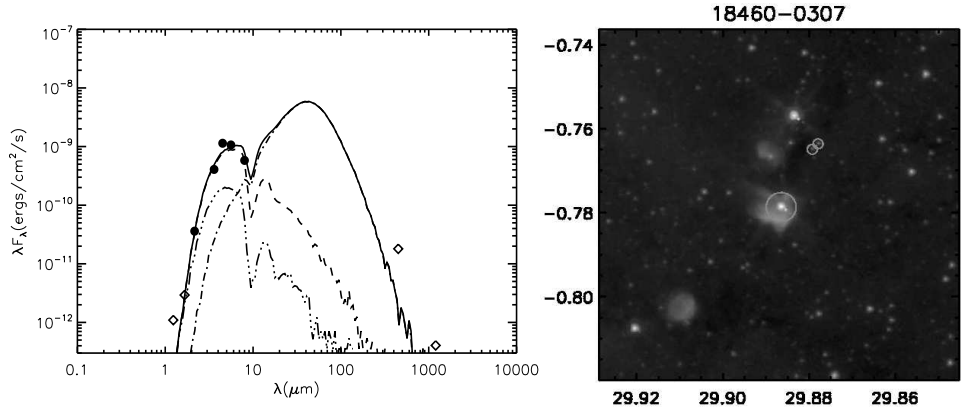} 
   \caption{The best fit SED model for the source IRAS18460-0307 and
an 8\mum\, grey scale image of the same. The solid line shows the best fit
model for an aperture of 15000AU, the dashed line represents disk
emission, dot-dashed line the envelope component and the
multiple-dot-dashed line the scattered component of the emission. Filled circles are data points and diamond symbols are upper limits.}

  \end{figure}

\subsection{ The SED fitting tool}

The SED fitting tool described by \citet{rob07} is based on a grid of
radiative transfer models described by RWIWD06. The models assume
axisymmetric young stellar objects with a photosphere, disk, and
infalling envelope. The upper limit of the mass range is 50\msun and
the physics that governs the higher mass regime is based on accretion
models described by \citet{whit04}. A pre-calculated grid of model
SEDs, computed using 14 variables are compared with the observed
data. The models for which the difference between the $\chi^2$ value
per data point of the fit and the best $\chi^2$ value per data point
is smaller than three are returned by the tool as the best fit models.
Multiple models are returned by the tool for a given set of data
points depending on the data quality, such as number of data points
and their errors that define the SED. This results in a degeneracy
which is typically 200 for the input SEDs we have used. In the
presence of such high degeneracy, we adopt a method of constructing
histograms for each physical parameter and finding the peak or by
plotting a two parameter plot for all models and choosing the median
of the distribution. For example to choose the best value of mass and
time, we would plot a two parameter plot of mass vs time given by all
the models, and if it results in a distribution centered on a
particular mass and time value, those values are taken as the median
representative values for mass and time of the source. These methods
are explained in detail by \citet{grave08}.

\subsection{ SED fitting results}

Figure.~2 shows the best fit model for the source IRAS18460-0307 along
with an 8\mum\, image of the source. The solid line shows the best fit
model for an aperture of 15000AU, the dashed line represents disk
emission, dot-dashed line the envelope component and the
multiple-dot-dashed line the scattered component of the emission. The
median values of the physical parameters from model fits, such as
mass, disk accretion rate, envelope accretion rate, stellar radii and
age were tabulated for all the 50 sources. In those cases where higher
spatial resolution data from the interferometers was available for the
millimeter or submm bands, the degeneracy of the model fits was very
low, demonstrating the importance of such observations at longer
wavelengths. Infact, a single model fit was obtained for the source
IRAS18089-1732.

The estimated physical parameters from the model fits span a range
consistent with the known physics of massive stars. These parameters
have a range of M = 6--45 \msun, R = 10-100 R$_{\odot}$, t =
10$^3$--10$^5$ yr, M$_{disk}$ = 0--0.9 \msun, and $\dot{M}_{env}\sim$
10$^{-4}$\msun yr$^{-1}$--10$^{-2}$\msun yr$^{-1}$. These values are
consistent with the input physics assumed in the SED model fitter.
While the observed data is well modelled by assumed accretion scenario
model, it is necessary to examine which of the estimated parameters
are independent of the limitations of the model. In Fig.~3 we plot the
accretion rate versus age of the source, representing the mass of the
sources using symbols of different sizes. Circles represents disk
accretion rate whereas triangles and squares represent envelope
accretion rate. Squares are sources without disks. It can be seen from
the figure that the most massive stars (biggest symbols, M$gt$30\msun)
are all squares, meaning that they do not pocess any disks and are
also among the youngest sources in the plot. The remaining symbols are
more or less evenly spread around in time but not in
$\dot{M}$. Although the model grid consists of many models for massive
stars with disks and envelopes together, the observed data best
matches with massive star models without disks. Therefore, the result
that ``envelope accretion is the dominant factor in building massive
stars'' and the value $\dot{M}_{env} \sim$ 10$^{-4}$--10$^{-2}$\msun
yr$^{-1}$ is independent of any limitations that could be inherent in
the model grid.

\begin{figure} 
  
  \includegraphics[width=11.5cm]{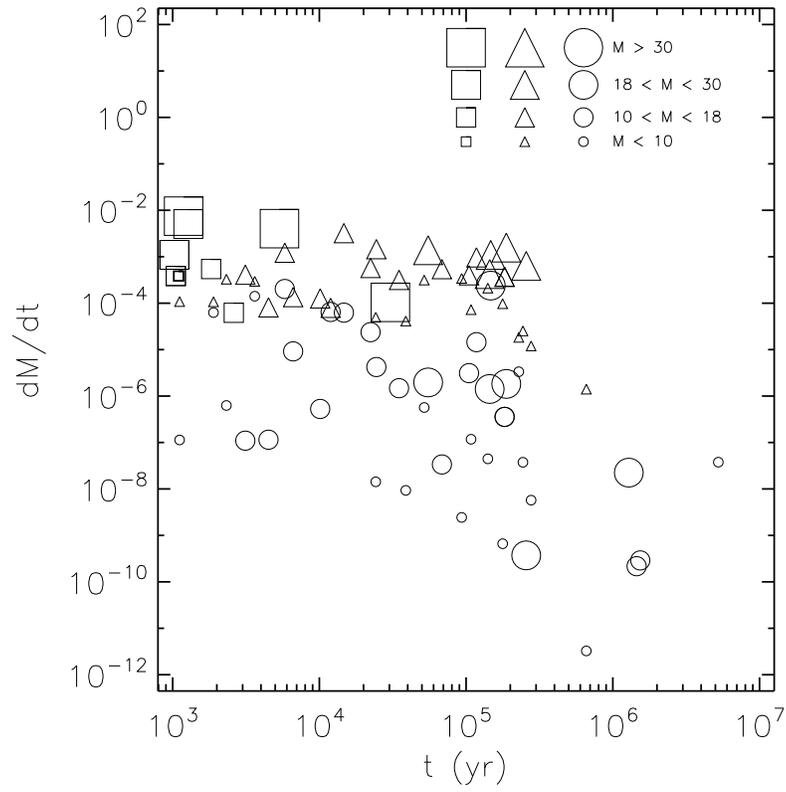} 

\caption{ 
Accretion rate versus age plot from the best fit models of 50
sources. Circles represents disk accretion rate whereas triangles and
squares represent envelope accretion rate. Sizes of the symbols are
proportional to the mass.}
\vskip -0.5cm
  \end{figure}

\section{Implications}

The SEDs of the IR counterparts to HMPO candidates can be well fitted
by accretion models involving an infalling envelope, disk,
photosphere, bipolar cavity and a radiative equilibrium solution. In
this scenario, the HMPO IRCs imitate the classical YSO phases known
from low mass star formation. The envelope, rather than the disk,
appears to be the major reservoir feeding the central engine in the
most massive young stars. Therefore probing the inner few thousand AU
structure of such sources are indispensable in understanding the
details of massive star formation \citep[e.g.][]{grave07}. 27 sources
with IR counterparts also have some level of unresolved centimeter
continuum emission which may represent the ionised emission arising
close to the central engines either in the form of expanding or
infalling ionised regions and/or from winds. The presence of ionised
gas together with an accretion set-up involving star/disk/envelope,
with high accretion rates, suggest the possibility of the ionised gas
in inward motions or forming a stable zone surrounding the central
engine. Therefore it appears that the theoretical scenarios involving
both ionised and molecular inflows \citep{keto02,keto03} may well be
the dominant mechanism producing the most massive stars.

\subsubsection{Reflection}

As evidenced by this meeting, there is an immediate need to better
define the terminology such as cores, protostars and evolutionary
stages in the massive star formation process. In the light of the
results presented here, it appears that the envelopes around massive
young stars are much larger than evidenced in the low mass star
formation and plays an important role as material resorvoir to feed
the central engine. The even larger flattenned structures/ toroids
observed in the millimeter regime, with sizes of 0.05 to 0.1pc may
represent the basic units which collapse to form the massive
stars. Thus, the definition of a {\em Core} may be better suited for
such flattenned structures/toroids.  We propose an evolutionary
sequence definition as follows: Stage I objects: millimeter bright
without an IR point source (e.g: IRDC's and mm only cores); Stage II:
IR bright (mm, IR, maybe weak cm continuum as well) with dominant
envelope accretion, central engine is not directly contributing to the
observed SED, high accretion rates and almost absent disks (sources
from this study); Stage III: IR sources with lower accretion rates,
near-infrared emission lines, central engine shows up in the
near-infrared (e.g: sources from the \citet{bik05} sample). Finally,
we note that high angular observations of massive outflows needs to
focus on investigating the ionised component and launching zones
because the outflows from massive young stars may arise from the
larger envelopes rather than disks according to the results of this
study.

\acknowledgements 

We greatly appreciate the efforts of Henrik Beuther in organising an
excellent conference and the SOC for providing the oppurtunity to
present this work. Kumar and Grave are supported by a research grant
PTDC/CTE-AST/65971/2006 approved by the FCT (The portuguese national
science foundation).


\newpage
\begin{thebibliography}{}

\bibitem[\protect\citeauthoryear{Beltr\'{a}n et al.}{2006}]{bel05} Beltr\'{a}n M. T., Cesaroni, R., Neri, R., Codella, C., Furuya, R. S., Testi, L., Olmi, L. 2005, A\&A, 435, 901
\bibitem[\protect\citeauthoryear{Beltr\'{a}n et al.}{2006}]{bel06} Beltr\'{a}n M. T., Brand, J., Cesaroni, R., Fontani, F., Pezzuto, S., Testi, L., Molinari, S. 2006, A\&A, 447, 221
\bibitem[\protect\citeauthoryear{Beuther et al.}{2002}]{beu02}Beuther, H., Schilke, P., Menten, K. M., Motte, F., Sridharan, T. K., Wyrowski, F. 2002, ApJ, 566, 945
\bibitem[\protect\citeauthoryear{Beuther et al.}{2007}]{beu07}Beuther, H., Leurini, S., Schilke, P., Wyrowski, F., Menten, K. M., Zhang, Q. 2007, A\&A, 466, 1065
\bibitem[\protect\citeauthoryear{Bik et al.}{2005}]{bik05}Bik, A., Kaper, L., Hanson, M. M., Smits, M., 2005, A\&A, 440,121
\bibitem[\protect\citeauthoryear{Churchwell et al.}{2006}]{churchwell06} Churchwell, E., Povich, M. S., Allen, D., Taylor, M. G., Meade, M. R. et al. 2006, ApJ, 649, 759
\bibitem[\protect\citeauthoryear{Fa\'{u}ndez et al.}{2004}]{faun04} Fa\'{u}ndez, S., Bronfman, L., Garay, G., Chini, R., Nyman, L.-\o{A}, May, J. 2004, A\&A, 426, 97 (Faun04)
\bibitem[\protect\citeauthoryear{Fontani et al.}{2002}]{fon02} Fontani, F., Cesaroni, R., Caselli, P., Olmi, L., 2002, A\&A, 389, 603 (Fon02)
\bibitem[\protect\citeauthoryear{Grave \& Kumar}{2007}]{grave07} Grave, J. M. C., Kumar, M. S. N., 2007, A\&A, 462, L37
\bibitem[\protect\citeauthoryear{Grave \& Kumar}{2008}]{grave08} Grave, J. M. C., Kumar, M. S. N., 2008, To be submitted to A\&A (Paper II)
\bibitem[\protect\citeauthoryear{Heitsch et al.}{2007}]{heitsch07} Heitsch, F., Whitney, B. A., Indebetouw, R., Meade, M. R., Babler, B. L, Churchwell, Ed. 2007, ApJ, 656, 227
\bibitem[\protect\citeauthoryear{Keto}{2002}]{keto02} Keto, E. 2002, ApJ, 580, 980
\bibitem[\protect\citeauthoryear{Keto}{2003}]{keto03} Keto, E. 2003, ApJ, 599, 1196
\bibitem[\protect\citeauthoryear{Kumar et al.}{2006}]{kumar06} Kumar, M. S. N., Keto, E. R., Clerkin, E. 2006, A\&A, 449, 1033
\bibitem[\protect\citeauthoryear{Kumar \& Grave}{2007}]{kumar07a} Kumar, M. S. N., Grave, J. M. C., 2007, A\&A, 472, 155 (Paper I)
\bibitem[\protect\citeauthoryear{Kurtz et al.}{1994}]{kurtz94} Kurtz, S. E., Churchwell, E., Wood, D. O. S., 1994, ApJS, 91, 659 
\bibitem[\protect\citeauthoryear{Molinari et al.}{1996}]{mol96} Molinari, S., Brand, J., Cesaroni, R., Palla, F. 1996, A\&A, 308, 573 (Mol96)
\bibitem[\protect\citeauthoryear{Molinari et al.}{1998}]{mol98} Molinari, S., Brand, J., Cesaroni, R., Palla, F., Palumbo, G. G. C. 1998, A\&A, 336, 339
\bibitem[\protect\citeauthoryear{Robitaille et al.}{2006}]{robitaille06} Robitaille, T. P., Whitney, B. A., Indebetouw, R., Wood, K., Denzmore, P., 2006, ApJSS, 167, 256  [RWIWD06]
\bibitem[Robitaille et al.(2007)]{rob07} Robitaille, T.~P., Whitney, B.~A., Indebetouw, R., \& Wood, K.\ 2007, \apjs, 169, 328 
\bibitem[\protect\citeauthoryear{Sridharan et al.}{2002}]{sri02}Sridharan, T. K., Beuther, H., Schilke, P., Menten, K. M., Wyrowski, F. 2002, ApJ, 566, 931 (Sri02)
\bibitem[\protect\citeauthoryear{Whitney et al.}{2004}]{whit04} Whitney, B. A., Indebetouw, R., Bjorkman, J. E., Wood, K., 2004, ApJ, 617, 1177
\bibitem[\protect\citeauthoryear{Whitney}{2005}]{whit05} Whitney, B. A., 2005, Nature, 437, 37
\bibitem[\protect\citeauthoryear{Williams et al.}{2004}]{wil04} Williams, S.~J., Fuller, G.~A., \& Sridharan, T.~K.\ 2004, \aap, 417, 115 
\bibitem[\protect\citeauthoryear{Zinnecker \& Yorke}{2007}]{zy07} Zinnecker, H., Yorke, H., 2007, ARA\&A, 45, 481
\end{thebibliography}
\end{document}